\documentstyle[preprint,aps]{revtex}

\begin{document}

\draft

\title{Astrophysical Bounds on Global Strings}

\author{Shane L. Larson
	\thanks{Electronic Address: shane@orion.physics.montana.edu}
	and William A.\ Hiscock
	\thanks{Electronic Address: billh@orion.physics.montana.edu}}

\address{Department of Physics, Montana State University, Bozeman,
Montana 59717}

\date{\today}
\preprint{MSUPHY97.08}
\maketitle

\begin{abstract}

Global topological defects produce nonzero stress-energy throughout
spacetime, and as a result can have observable gravitational
influence on surrounding matter.  Gravitational effects of global
strings are used to place bounds on their cosmic abundance.  The
minimum separation between global strings is estimated by considering
the defects' contribution to the cosmological energy density.  More
rigorous constraints on the abundance of global strings are
constructed by examining the tidal forces such defects will have on
observable astrophysical systems.  The small number of observed
tidally disrupted systems indicates there can be very few of these
objects in the observable universe.

\end{abstract}

\pacs{ }

\section{INTRODUCTION}

Topological defects in a spacetime arise from the spontaneous breaking
of local and global gauge symmetries.  Such defects may have been
created during vacuum phase transitions in the early universe, and
could have interesting cosmological implications, particularly in the
production of density fluctuations leading to the formation of large
scale structures \cite{VilRev,VilEv}.

Defects associated with local symmetry breaking, such as gauged
(magnetic) monopoles and gauged cosmic strings, are known to have no
substantial stress-energy associated with the fields outside the core
of the defect.  As a result, the spacetime exterior to the defect is a
vacuum solution of the Einstein equations\cite{Pres,Greg}, and, in the
case of the gauged cosmic string, it is actually
flat\cite{Gott,HisStr}.  Individual gauged monopoles have
insignificant gravitational effects, while gauged cosmic strings are
influential only when radiation or matter passes on opposite sides of
the string ({\it e.g.}, gravitational lensing).  Domain walls have
large gravitational effects \cite{IpSik}, which may be used to
strongly constrain their existence in the early universe.

Global defects differ from gauge defects in that there is no gauge
field associated with the symmetry breaking.  As a result, the
scalar fields vary spatially throughout the spacetime, yielding a
nonzero stress-energy tensor which can lead to significant
gravitational effects \cite{BarVil,CoKap,GiOrRu}.

Previous work \cite{HisMon} has shown that the gravitational
fields associated with a single global monopole can be used to
constrain the number density of such objects in the observable
universe. In this paper, similar considerations are applied to
the spacetime of a global string to develop astrophysical constraints
which will provide upper bounds to the abundance of such defects.
While the constraints developed here are based on the spacetime
metric of an infinitely long, cylindrically symmetric global string,
they may also be applied to string loops, so long as the distance to
the string is less than the local radius of curvature of the loop.

In Section II the exact metric describing a static, cylindrically
symmetric global string \cite{CoKap} is reviewed and an approximation
to the metric is developed which will allow the constraints to be
stated in analytic form.

In Section III bounds are constructed based on the strings'
contribution to the total cosmological energy density. Based on an
extremely simple model, the average distance between
global strings today must exceed $30\:[\eta\,/\,(10^{16}\:\mbox{GeV})\,]
\;\mbox{Mpc}$, where $\eta$ is the energy scale of the strings.

In Section IV somewhat more general bounds are developed by examining
the possible tidal disruption of various astrophysical systems by the
gravitational field of a global string.  A general expression is
developed for the tidal disruption distance, $d$, which is the closest
a system bound by gravity can approach a global string without being
tidally disrupted by the string's spacetime curvature.  Several
example constraints are developed using this expression.  In order to
prevent global strings from tidally disrupting an unacceptable
fraction of all galaxies, it is shown that strings must be separated by a
distance exceeding $3\:[\eta\,/\,(10^{16}\:\mbox{GeV})\,]\;\mbox{Mpc}$.  A
bound at lower string energy scales may be obtained by considering the
Earth-Moon system.  Lunar laser ranging data assures that no global
string with an energy scale  $\eta >7 \times 10^8 \;\mbox{GeV}$ has passed
within 5 times the distance between the Earth and Moon within the past
28 years.

Sign conventions and notation follow those of Misner, Thorne,
and Wheeler \cite{MTW}; natural units ($G = c = \hbar = 1$) are used 
throughout.

\section{SPACETIME OF A GLOBAL STRING}

The simplest example of a scalar field Lagrangian density for which
global strings are a possible consequence of symmetry breaking is a
single complex scalar field with a global U(1) symmetry \cite{VilRev}:
\begin{equation}
      {\cal L} = \partial_\mu \phi^\dagger \partial^\mu \phi - {1 \over 2}
        \lambda\left( \phi^\dagger \phi - \eta^2 \right)^2  .
	\label{Lagrangian}
\end{equation}
This Lagrangian is invariant under the global gauge transformation   
\begin{equation}
	\phi \rightarrow \tilde \phi = e^{i\Lambda}\phi  .
	\label{Gauge}
\end{equation}
where $\Lambda$ is an arbitrary constant.
A global string solution has the general form
\begin{equation}
	\phi = F(r)e^{i\theta}  ,
	\label{gstring}
\end{equation}
with $F(0) = 0$ and $F(r) = \eta$ for all $r \ge r_C$, where $r_C$
is the core radius of the string, roughly given by $r_C \approx 1/
\eta$. In a typical grand unified theory, the parameter $\eta$ might
be of order $10^{16}$ GeV.

Cohen and Kaplan \cite{CoKap}  have found the exact spacetime metric
for a static, cylindrically symmetric global string such as that
described above. The metric is 
\begin{equation}
	ds^2 = \left( {u \over u_0}  \right) \left( - dt^2 + dz^2
        \right) + A(u) \left( du^2 + d\theta^2 \right),
	\label{Metric}
\end{equation}
where
\begin{equation}
	 A(u) = {1 \over \eta^2} \sqrt{u_0 \over u} \exp \left[
         (u_0^2 - u^2)/u_0 \right],
	  \label{Guu}
\end{equation}
and
\begin{equation}
	 u_0 = {1 \over {8 \pi \eta^2}}.
	  \label{U0}
\end{equation}

Here $u$ is a dimensionless radial coordinate that decreases outward
from the string.  The axis of cylindrical symmetry is located at
$u = \infty$, the surface of the string is at $u \simeq u_0$, and
there is a timelike curvature singularity at $u = 0$, which is at
a finite proper spatial distance from the core. The metric given
in Eq.(\ref{Metric}) is valid outside the core region, $ u \lesssim
u_0$.

Cohen and Kaplan have derived an elementary upper bound on the energy
scale of the global string by requiring that the distance from the
core to the outer singularity be greater than the size of the present
cosmological horizon.  This places an upper limit of $\eta \lesssim
2.0 \times 10^{17}$ GeV on the energy scale of the string.

The nature of the dimensionless radial coordinate $u$ makes intuition about 
the global string system difficult.  In order to gain further insight into 
the global string spacetime, it is desirable to devise a method by which 
proper radial distances from the string can be obtained for given values 
of $u$.  One approach is to directly calculate the proper radial distance 
from the metric, integrating along a curve of fixed $t$, $\theta$, and $z$ 
from the surface of the string at $u_0$ to the desired location at $u$:
\begin{equation}
    r_p = \int ds = \int_u^{u_0} \sqrt{g_{uu}} du = \int_u^{u_0}
    {1 \over \eta}\left( u_0 \over u \right)^{1 \over 4} \exp
    \left[{u_0^2 - u^2} \over {2u_0} \right] du.
    \label{RInteg}
\end{equation}
This has the disadvantage that it does not appear to be possible to 
evaluate the integral in terms of simple functions.  A much simpler method 
is to use the cylindrical symmetry of the system to advantage by defining a 
circumferential radius (analogous to the usual area or curvature coordinate 
$r$ in the Schwarzschild metric) by ${r_c}^2 = g_{\theta\theta}$, or
\begin{equation}
    r_c = \sqrt{g_{\theta\theta}} = {1 \over \eta} \left( u_0 \over
    u \right)^{1 \over 4} \exp \left[{u_0^2 - u^2} \over {2 u_0}
    \right].
    \label{RCirc}
\end{equation}
One may use these expressions for $r_p$ and $r_c$ to determine what
values of $u$ correspond to particular distances for a variety of
energy scales, $\eta$, for the string. Astrophysically relevant
distance scales turn out to generally correspond to values of $u$
close to $u_0$.

An approximation scheme can be developed to simplify
computations in this spacetime by expanding the metric functions in
power series for values of $u$ close to $u_0$. Defining a new
coordinate $\delta$ by
\begin{equation}
     \delta = {{u_0 - u} \over u_0} ;
     \label{ApproxU}
\end{equation}
the metric, rewritten in terms of the new coordinate and expanded to
first order in $\delta$, becomes
\begin{equation}
        ds^2 = (1-\delta)(-dt^2+dz^2) + {1 \over \eta^2}
        e^{2u_0 \delta}({u_0}^2 d\delta^2 + d\theta^2) .
        \label{ApproxA}
\end{equation}
At first glance, one might object to this form of the metric on the
grounds that the approximation does not appear to have been carried
out to completion.  The appearance of the exponential function
containing the small parameter $\delta$ suggests that the exponential
should be expanded to first order in $\delta$.  This is not possible
because the argument of the exponential is $u_0 \delta$, which in the
domains of physical interest typically has values of order $10^2$.
This approximation is similar, but not identical to, the linearized
global string metric discussed in Refs.\cite{CoKap,HarSik}.  

Using this approximation the proper and circumferential radii may be
simply expressed in terms of the coordinate $\delta$ as
\begin{equation}
     r_p = r_c = {e^{u_0 \delta}\over \eta} \equiv r  , 
     \label{ApproxRInt}
\end{equation}
where the last equality indicates that a new coordinate $r$ is defined by 
this equation, within the context of the approximate metric.

The approximate metric then takes on a particularly simple form when
expressed in terms of the coordinate $r$,
\begin{equation}
ds^2 \approx \left[1 - {1 \over u_0}\ln(r\eta)\right]\left(-dt^2
	+d\eta^2 \right) + dr^2 + r^2 d\theta^2 .
     \label{AMetricR}
\end{equation}
Written in this form, our approximate metric is seen to be equivalent to 
the linearized metric\cite{CoKap,HarSik}, with the further simplification 
of ignoring terms of order ${1/u_0}$ in $g_{\theta\theta}$, but not in 
$g_{tt}$ and $g_{zz}$.  In every case examined, we have checked the 
validity of the results by comparison with numerical integration using 
the full metric of Eq.(\ref{Metric}).  Results will generally be 
stated in terms of the approximate coordinate $r$, as this allows 
simple analytic forms for the constraints and simple interpretation of 
the radial coordinate.

\section{COSMOLOGICAL DENSITY CONSTRAINTS}

Global strings differ from cosmic strings in that they have
non-constant scalar fields extending throughout the spacetime, and
hence will have significant gravitational influence on observable
astrophysical systems.  This fact can be used to construct constraints
on the abundance of global strings at the current epoch.

The scalar fields associated with a global string will contribute to
the total energy density of the spacetime.  An average energy density
due to these scalar fields, $\bar{\rho}$, within a cylindrical volume,
can be estimated by dividing the integrated mass in the scalar field
by the cylindrical volume of the spacetime out to fixed radius $r$.
A number of different estimates may be constructed, depending upon
how one imagines the global strings filling spacetime within a
cosmological model. The simplest possible model is examined here,
assuming the universe to be filled with a network of parallel global
strings and treating each string as if it is in flat space
(appropriate to the approximate metric of Eq.(\ref{AMetricR}).
The energy density in the scalar fields is approximately given by
\begin{equation}
    \rho \sim {{\eta^2} \over {r^2}}   ,
     \label{DensityScalar}
\end{equation}
If the typical distance between strings is defined to be $2r_0$, then
the mass contributed by the scalar fields within the radius $r_0$, per
unit length, is
\begin{equation}
	M \sim 2 \pi \int_0^{r_0} {{\eta^2}\over{r^2}} r dr \sim 2 \pi 
	\eta^2 \ln(r_0 \eta)  .
\label{gsmass}
\end{equation}
This mass occupies the spatial volume per unit length of string 
\begin{equation}
	V \sim \pi {r_0}^2   ,
\label{gsvol}
\end{equation}
leading to an average mass density given by
\begin{equation}
     \bar{\rho} \sim {\eta^2 \over {r_0}^2} \ln(r_0 \eta)  .
     \label{Density}
\end{equation}

A constraint on the abundance of global strings may be obtained from
this result by insisting that $\bar{\rho}$ be less than ten times
the closure density of the universe, a rough upper limit on the total
cosmological density of matter consistent with present astronomical
observations. The separation for all energy scales $\eta$ of potential 
interest is such that the logarithmic term in Eq.(\ref{Density}) will
always be of order $10^2$. This implies a minimum separation of
\begin{equation}
	r_0 \ge 7 \times 10^{26}\:h^{-1}\: \left( {\eta} \over {10^{16}\:
	\mbox{GeV}} \right) \;\mbox{cm}\:\approx 20\:h^{-1}\:\left( {\eta} 
	\over {10^{16}\:\mbox{GeV}} \right)\;\mbox{Mpc} , 
     \label{DensitySep}
\end{equation}
where $h$ is the Hubble constant in units of $100\;{\rm km\:sec^{-1}\:
Mpc^{-1}}$.
This limit depends rather strongly on the model used for the string
distribution, the assumption that all strings are infinite in length
(no loops), etc.  More rigorous constraints can be placed on global
string abundances by considering limits which do not depend on
cosmological assumptions, but are instead determined by local
astrophysical effects of a single string.

\section{TIDAL ACCELERATION CONSTRAINTS}

The stress-energy of the global scalar fields associated with the
string will give rise to nonzero curvature of the spacetime.  Extended
astrophysical systems will experience tidal accelerations due to the
string; these accelerations may be examined by using the equation of
geodesic deviation.  Given an extended body whose center of mass
follows a geodesic with four-velocity $u^\alpha$, the acceleration
of a point in the the body relative to the center of mass is given by
\begin{equation}
     a^\mu = -R^\mu{}_{\alpha\beta\nu} u^\alpha n^\beta u^\nu
     \label{GeodesicDeviation}
\end{equation}
where $n^\alpha$ is the body vector orthogonal to $u^\alpha$
connecting the point to the center of mass, and
$R^\mu{}_{\alpha\beta\nu}$ is the Riemann tensor of the background
(global string) spacetime.

Astrophysical systems such as the solar system, the galaxy, etc.  are
bound by gravity.  Such systems may be tidally disrupted, ``ionized''
in a sense, if they should approach a global string too closely.  By
examining the tidal accelerations ($a_T$) caused by a global string
using Eq.(\ref{GeodesicDeviation}) and comparing them to the Newtonian
gravitational accelerations binding the system together ($a_N$), a
tidal disruption distance $d$ may be computed such that at the
distance $d$,
\begin{equation}
     a_{T} = a_{N}.
     \label{IonizeCondition}
\end{equation}
If a system passes closer to a global string than $d$, it will be
tidally disrupted.

It is convenient to compute the Riemann tensor components in an orthonormal 
frame; this simplifies the later calculation of the components of the tidal 
acceleration in the frame attached to a freely falling observer.  The basis 
one-forms for an orthonormal frame which is static with respect to the 
string may be expressed as
\begin{equation}
\omega^{\hat 0} = \left({u \over u_0} \right)^{1/2} dt \approx
        \left(1-{1 \over {2u_0}}\ln(r\eta)\right)dt ,
\label{omegat}
\end{equation}
\begin{equation}
\omega^{\hat 3} = \left({u \over u_0} \right)^{1/2} dz \approx
        \left(1-{1 \over {2u_0}}\ln(r\eta)\right)dz ,
\label{omegaz}
\end{equation}
\begin{equation}
\omega^{\hat 1} = {A(u)}^{1/2} du \approx
        dr,
\label{omegau}
\end{equation}
and
\begin{equation}
\omega^{\hat 2} = {A(u)}^{1/2} d\theta \approx
        rd\theta .
\label{omegath}
\end{equation}
In such a frame, the components of the Riemann tensor can be
written
\begin{equation}
     R^{\hat{2}\hat{0}}{}_{\hat{2}\hat{0}} = {1 \over A(u)} 
     \left[ {1 \over {8 u^2} } + {1 \over {2 u_0} } \right]
     \approx {1 \over {2u_0r^2}},
     \label{Rotot}
\end{equation}
\begin{equation}
     R^{\hat{2}\hat{1}}{}_{\hat{2}\hat{1}} = {1 \over
     {2 A(u)}}\left[ {2 \over {u_0} } - {1 \over {2 u^2} } \right]
     \approx {1 \over {u_0r^2}} ,
     \label{Rouou}
\end{equation}
\begin{equation}
     R^{\hat{3}\hat{0}}{}_{\hat{3}\hat{0}} = -{1 \over {4 A(u)}} {1
     \over u^2} \approx {-1 \over {4{u_0}^2r^2}} ,
     \label{Rztzt}
\end{equation}
\begin{equation}
     R^{\hat{3}\hat{1}}{}_{\hat{3}\hat{1}} = {1 \over A(u)} 
     \left[ {1 \over {8 u^2} } - {1 \over {2 u_0} } \right]
     \approx {-1 \over {2u_0r^2}}  ,
     \label{Rzuzu}
\end{equation}
\begin{equation}
     R^{\hat{0}\hat{1}}{}_{\hat{0}\hat{1}} = {1 \over A(u)} 
     \left[ {1 \over {8 u^2} } - {1 \over {2 u_0} } \right]
     \approx {-1 \over {2u_0r^2}},
     \label{Rtutu}
\end{equation}
where only the lowest order terms in ${1/u_0}$ have been kept in the 
approximate expressions.  The components of the tidal acceleration felt by 
an extended body are given by Eq.(\ref {GeodesicDeviation}), and are 
computed in an orthonormal frame attached to a radially freely falling 
observer.  The four-velocity of such an observer is
\begin{equation}
u^{\hat 0} = \gamma ,
\label{ut}
\end{equation}
\begin{equation}
u^{\hat 1} = \gamma v ,
\label{uu}
\end{equation}
where $\gamma = 1/\sqrt{1-v^2}$, and $v$ is the radial three-velocity
measured by a static observer. The freely falling frame is related
to the static orthonormal frame by a simple Lorentz transformation.
The orthonormal basis vectors of the freely falling frame, 
$e_{\tilde \alpha}$, have components in the static orthonormal frame
given by: 
\begin{equation}
{e_{\tilde 0}}^{\hat \alpha} = u^{\hat \alpha},\;{e_{\tilde 1}}^{\hat
        \alpha} = n^{\hat \alpha},\;{e_{\tilde 2}}^{\hat \alpha} =
        {\delta_{2}}^{\hat \alpha},\;{e_{\tilde 3}}^{\hat \alpha} =
        {\delta_{3}}^{\hat \alpha},
\label{ffframe}
\end{equation}
where $n^{\hat \alpha}$ is the unit length spacelike vector in the
$\hat 0, \hat 1$ (or $(t,u)$) plane orthogonal to $u^\alpha$.
Computing the components of the tidal accelerations in the freely
falling frame then yields
\begin{equation}
     a_{\tilde 1} = \ell_{\tilde 1} R^{\hat{0}\hat{1}}{}_{\hat{0}
     \hat{1}} ,
     \label{Acceln}
\end{equation}
\begin{equation}
    a_{\tilde 2} = \ell_{\tilde 2} \gamma^2 \left[ 
    R^{\hat{2}\hat{0}}{}_{\hat{2}\hat{0}}
    - v^2 R^{\hat{2}\hat{1}}{}_{\hat{2}\hat{1}} \right] ,
    \label{Accelth}
\end{equation}
\begin{equation}
    a_{\tilde 3} = \ell_{\tilde 3} \gamma^2 \left[
    R^{\hat{3}\hat{0}}{}_{\hat{3}\hat{0}}
    - v^2 R^{\hat{3}\hat{1}}{}_{\hat{3}\hat{1}}\right] ,
    \label{Accelz}
\end{equation}
where $\ell^{\tilde \alpha}$ is the body vector of the system. 
Substituting the approximate values for the Riemann tensor
components from Eqs.(\ref{Rotot}-\ref{Rtutu}),
\begin{equation}
a_{\tilde 1} \approx {-\ell_{\tilde 1} \over {2u_0r^2}}  ,
\label{a1}
\end{equation}
\begin{equation}
a_{\tilde 2} \approx {{\ell_{\tilde 2}\gamma^2} \over {2u_0r^2}}
        \left(1-2v^2\right)  ,
\label{a2}
\end{equation}
\begin{equation}
a_{\tilde 3} \approx {{\ell_{\tilde 3}\gamma^2} \over {4{u_0}^2r^2}}
        \left(2u_0v^2 - 1 \right)  .
\label{a3}
\end{equation}

The maximum tidal acceleration on an extended system depends on the 
orientation of its body vectors and its velocity relative to the string.  
Global strings are, from a Newtonian viewpoint, gravitationally 
repulsive\cite{CoKap}; as seen by static observers, a body falling inward 
on a radial geodesic slows, reaches a turning point at a minimum radius, 
and then accelerates (in a coordinate sense) outward from the string.  
Since the tidal accelerations depend inversely on the distance to the 
string, but directly on the velocity relative to the string, it is not 
clear {\it a priori} where the tidal accelerations will be maximized.  
Examination of the tidal accelerations for the exact Riemann tensor and 
metric shows that the tidal accelerations diverge as $u \rightarrow 0$, as 
expected, as that is the curvature singularity surrounding the string.  
Once away from the curvature singularity, the tidal accelerations pass 
through a local maximum at the point of closest approach to the string for 
the cases of interest.  The strongest constraints on astrophysical 
distances to global strings will therefore be obtained by setting $ v = 0, 
\gamma =1$ in Eqs.  (\ref{a1}-\ref{a3}).  Examination of 
Eqs.(\ref{a1}-\ref{a3}) shows that the tidal accelerations will generically 
be of order
\begin{equation}
a_T \approx {{\ell} \over {2u_0r^2}} .
\label{ageneric}
\end{equation}
If we then set this acceleration equal to the Newtonian acceleration
binding a system of size $\ell$ and mass $M$ together, the critical 
tidal disruption distance is (using Eq.(\ref{U0})):
\begin{equation}
     d \simeq 2 \,\sqrt{\pi} \,\eta\, {\ell^{3/2} \over 
     M^{1/2}} .
     \label{critd}
\end{equation}
This approximate equation has been compared with exact results obtained
by numerical integration of the geodesic deviation equation, utilizing
the exact metric of Eq.(\ref{Metric}). The approximation is found to
be remarkably good for all astrophysical systems of interest,
agreeing with the exact results to about one tenth of a percent or better.
Above $10^{17}$ GeV, the approximation breaks down due to the 
proximity of the curvature singularity at $u = 0$, which moves inward
as the energy scale is increased.

The expression of Eq.(\ref{critd}) may be used to construct constraints on 
the allowable distances between strings (lest all astrophysical systems be 
disrupted) or between specific systems and the nearest global string.  In 
the remainder of this section, the tidal disruption distance is evaluated 
for several systems of interest, and some basic constraints on how 
plentiful global strings may be are thereby developed.

The expression for the tidal disruption distance in Eq.(\ref
{critd}) is evaluated for several systems of interest below, with
conventional units restored:
\begin{equation}
     \mbox{Earth-Moon} \quad
     d \simeq 850  \left({\eta \over {10^{16}\:\mbox{GeV}}} \right)
     \ell_{\rm Earth-Moon} = 3.3 \times 10^8 \ \left( \eta \over
     {10^{16}\:\mbox{GeV}} \right)\   \mbox{km} ,
     \label{IonizeMoon}
\end{equation}

\begin{equation}
     \mbox{Sun-Pluto} \quad
     d \simeq 180  \left(\eta \over {10^{16}\:\mbox{GeV}} \right)
     \ell_{\rm Pluto} = 7300 \ \left( \eta \over
     {10^{16}\:\mbox{GeV}} \right)\  \mbox{AU}   ,
     \label{IonizePluto}
\end{equation}

\begin{equation}
     \mbox{Galaxy} \quad
     d \simeq 5.2  \left(\eta \over {10^{16}\:\mbox{GeV}} \right)
     l_{\rm galaxy} \simeq 150 \  \left( \eta \over
     {10^{16}\:\mbox{GeV}} \right)\mbox{kpc} ,
     \label{IonizeGalaxy}
\end{equation}

\begin{equation}
     \mbox{Cluster of Galaxies} \quad
     d \simeq 5.2 \left(\eta \over {10^{16}\:\mbox{GeV}} \right)
     l_{\rm cluster} \simeq 15  \left( \eta \over
     {10^{16}\:\mbox{GeV}} \right)\  \mbox{Mpc} ,
     \label{IonizeCluster}
\end{equation}
where $\ell_{\rm Earth-Moon}$ and $\ell_{\rm Pluto}$ are the semi-major 
axes of the orbits of those systems, and $\ell_{\rm galaxy}$ and 
$\ell_{\rm cluster}$ are the ``typical'' radii of those systems.  
These constraints, and similar bounds constructed for other systems, 
may be used to place bounds on the distance between global strings, 
or between the Earth and the nearest string.

The use of the equation of geodesic deviation is only valid if the extended 
body is small compared to its separation from the string.  The critical 
distances listed above based on disruption of a galaxy or a cluster are 
then only valid if the string energy scale is $\gtrsim 10^{16} \mbox{GeV}$; 
for the smaller systems (Earth-Moon and Sun-Pluto), the derivation is valid 
for an order of magnitude or so smaller energy.  If one considers a system 
which experiences tidal perturbations smaller than total disruption, then 
the critical distance is increased, and the minimum string energy 
constrained is reduced.

As an example, consider the Earth-Moon system.  The Moon's orbit has been 
studied for centuries, with steadily increasing precision, until in the 
last 28 years lunar laser ranging\cite{Nord} has reduced the uncertainty in 
the Earth-Moon distance to $< 3\ \mbox{cm}$.  If one assumes that the 
Earth-Moon system has not suffered an unexplained acceleration (perhaps due 
to a passing global string) sufficient to cause a 3 cm change in the 
Earth-Moon distance, then this implies that the global string tidal 
acceleration is bounded above by
\begin{equation}
a_{T} \lesssim {M \over \ell_{\rm Earth-Moon}^2}-{M \over
        {\left(\ell_{\rm Earth-Moon}+3\:\mbox{cm}\right)^2}} \approx
        1.6 \times 10^{-10} a_{N}
\label{laser}
\end{equation}
The distance $d$ within which no string must have passed, lest it create
such a perturbing acceleration, may then be evaluated from Eq.(\ref{laser})
using Eq.(\ref{ageneric}) for the tidal acceleration
\begin{equation}
d_{llr} \gtrsim 0.84\,\left(\eta \over {10^{16}\:\mbox{GeV}}
        \right)\;\mbox{pc}.
\label{llr}
\end{equation}
One can then state that, on the basis of lunar laser ranging,
no global string with an energy scale of
$10^{16}$ GeV has passed within a parsec of the Earth-Moon system
within the last 28 years. Alternately, examining the lower-energy
limit of this bound, if we demand that $d > 5 \ell$ for the geodesic
deviation equation to be valid, then Eq.(\ref{llr}) implies that no
global string with an energy scale  $\eta > 7 \times 10^8$ GeV has passed
within 5 times the distance between the Earth and Moon within the past
28 years.

Another approach to constructing limits may be illustrated by
considering the tidal disruption of galaxies.  It has been found that
roughly 10\% of all observed galaxies exhibit evidence of tidal
disruption \cite{Vor}.  Most of these tidal events are clearly due to
galaxy-galaxy interactions; at most perhaps 10\% of these events (or 
1\% of all galaxies) could possibly be disrupted due to close passage of a
global string.  If one insists that global strings be separated
sufficiently so that at most 1\% of all galaxies will pass within a
distance $d$, given by Eq.(\ref{IonizeGalaxy}), of a string, then that
implies that the minimum separation of global strings today is
\begin{equation}
s \gtrsim 20\:d \approx 3 \ \left( \eta \over
     {10^{16}\:\mbox{GeV}} \right)\;\mbox{Mpc} .
\label{statgal}
\end{equation}
While the constraint in Eq.(\ref{statgal}) is weaker than the
cosmological density limit of Eq.(\ref{DensitySep}), it is
considerably more robust, since it is based on direct astronomical
observations, and is independent of assumptions about cosmological
models and the global distribution of strings.

\acknowledgements
The work of W.\ A.\ H. was supported in part by National Science
Foundation Grant No. PHY-9511794.

\end{document}